\documentclass[graybox]{svmult}

\usepackage{mathptmx}       
\usepackage{helvet}         
\usepackage{courier}        
\usepackage{type1cm}        
\usepackage{makeidx}         
\usepackage{graphicx}        
\usepackage{multicol}        
\usepackage[bottom]{footmisc}

\makeindex             

\begin{document}

\title*{Formation, evolution and multiplicity of brown dwarfs and giant
exoplanets}
\titlerunning{Formation, evolution and multiplicity of brown dwarfs and giant
exoplanets}
\author{Jos\'e~A. Caballero}
\institute{Jos\'e~A. Caballero \at Departamento de Astrof\'{\i}sica y
Ciencias de la Atm\'osfera, Facultad de F\'{\i}sica, Universidad Complutense de
Madrid, E-28040 Madrid, Spain,
\email{caballero@astrax.fis.ucm.es}}
%
\maketitle

\abstract*{This proceeding summarises the talk of the awardee of the Spanish
Astronomical Society award to the the best Spanish thesis in Astronomy and
Astrophysics  in the two-year period 2006--2007.
The thesis required a tremendous observational effort and covered many different
topics related to brown dwarfs and exoplanets, such as the study of the mass
function in the substellar domain of the young $\sigma$~Orionis cluster down to
a few Jupiter masses, the relation between the cluster stellar and substellar
populations, the accretion discs in cluster brown dwarfs, the frequency of very
low-mass companions to nearby young stars at intermediate and wide separations,
or the detectability of Earth-like planets in habitable zones 
around ultracool (L- and T-type) dwarfs in the solar neighbourhood.}

\abstract{This proceeding summarises the talk of the awardee of the Spanish
Astronomical Society award to the the best Spanish thesis in Astronomy and
Astrophysics  in the two-year period 2006--2007.
The thesis required a tremendous observational effort and covered many different
topics related to brown dwarfs and exoplanets, such as the study of the mass
function in the substellar domain of the young $\sigma$~Orionis cluster down to
a few Jupiter masses, the relation between the cluster stellar and substellar
populations, the accretion discs in cluster brown dwarfs, the frequency of very
low-mass companions to nearby young stars at intermediate and wide separations,
or the detectability of Earth-like planets in habitable zones 
around ultracool (L- and T-type) dwarfs in the solar neighbourhood.} 

\bigskip
{\flushright {\em El que ama arde \\ y el que arde vuela a la velocidad de la
luz}\\ Lagartija Nick (Val del Omar)\\}

\section{``De fuscis pusillis astris et giganteis exoplanetis'' (Part I)} 
\label{sec:part1}


The recipient of the Spanish Astronomical Society ({\em Sociedad Espa\~nola de
Astronom\'{\i}a}) award to the best Spanish thesis in Astronomy and
Astrophysics in the two-year period 2006--2007 was the thesis
``Formation, evolution and multiplicity of brown dwarfs and giant exoplanets''
({\em ``Formaci\'on, evoluci\'on y multiplicidad de enanas marrones y
exoplanetas gigantes''}), by the author of this proceeding.
It was supervised by R. Rebolo and V.~J.~S. B\'ejar and defended at
the Universidad de La Laguna/Instituto de Astrof\'{\i}sica de Canarias in
March 2006.  

My thesis was an ambitious initiative to search for the answers to some key
questions in Astrophysics: 
{\em How and where do the substellar objects form? What are their properties?
How are they related to stars?}
Such answers should be obtained through observations at 1--10\,m-class
telescopes, especially in the red optical and the near-infrared.
Just to illustrate the amount and variety of data eventually collected or the
difficulty in summarising the thesis in a short talk or in this proceeding,
during my PhD, I~observed during 192 telescope nights with 18 different
instruments in 11 different telescopes, not counting data aquired by other
observers or with space missions (e.g. {\em Hubble}, {\em Spitzer}, {\em
XMM-Newton}). 
I~splitted the 459 pages of the thesis into five parts, 11 chapters and three
appendices, which can be downloaded from a public ftp 
site\footnote{\tt ftp://astrax.fis.ucm.es/pub/users/caballero/PhD.}.
Most of the chapters have been the basis of many refereed publications in main
international journals. 
The used language was Spanish.

\subsection{Brown dwarfs and objects beyond the deuterium-burning mass limit
(Chapter 1)} 
\label{sec:chapter1}

This was the necessary introductory chapter of the thesis.
It dealt with the following subjects:
\begin{itemize}
\item Physical properties of substellar objects:
basic definitions, hydrogen and deu\-te\-ri\-um-burning mass limits, lithium
test; time evolution of physical parameters (luminosity, temperature, absolute
magnitudes, colours);
ultracool atmospheres, new spectral types L and T, meteorology. 
\item An historical view of the searches of substellar objects (with an
interesting discussion on which was the first brown dwarf: Teide~1 --Rebolo
et~al. 1995, 1996--, GJ~229~B --Nakajima et~al. 1995--, PPL~15 --Stauffer et~al.
1994; Basri et~al. 1996--, HD~114762~b --Latham et~al. 1989--, GD~165B --Becklin
\& Zuckerman 1988; Kirkpatrick et~al. 1999-- or LP~944--20 --Luyten \& Kowal
1975; Tinney 1998--); 
\item Theoretical scenarios of formation of substellar objects and planetary
systems.
\item Young star clusters, photometric searches and the substellar initial mass
function. 
\item Ultracool companions to stars, multiplicity of L and T dwarfs,
circumsubestellar discs (with compilations of late-type companions and very
low-mass binaries).
\end{itemize}

The chapter ended with the main aims of the thesis, which were studying
the mass function in the substellar domain of the $\sigma$~Orionis cluster
($\tau \sim$ 3\,Ma) down to a few $M_{\rm Jup}$, the relation between the
cluster stellar and substellar populations, the accretion discs in young cluster
brown dwarfs, the frequency of very low-mass companions to nearby young stars
($\tau \sim$ 100\,Ma) at intermediate and wide separations, and the
detectability of Earth-like planets in habitable zones around ultracool (L and
T) dwarfs in the solar neighbourhood.

\section{The substellar population in $\sigma$~Orionis and its relation with the
stellar population (Part II)} 
\label{sec:part2}

\subsection{The $\sigma$~Orionis cluster (Chapter 2)} 
\label{sec:chapter2}

\begin{table}[b]
\caption{Main parameters of the $\sigma$~Orionis open cluster.}
\label{thetable}       
\begin{tabular}{l ccc l}
\hline\noalign{\smallskip}
Parameter 		& Canonical 		& Min.:max. 		& Unit		& Key references	\\
			& value			& values	 	& 		&		\\
\noalign{\smallskip}\svhline\noalign{\smallskip}
Age, $\tau$		& 3			& 1:8	 		& Ma		& Zapatero Osorio et al. 2002a; Sacco et~al. 2006	\\ 
Distance, $d$		& 385			& 330:470	 	& pc		& Mayne \& Naylor 2008; Caballero 2008b			\\ 
$E(B-V)$		& 0.07			& 0.00:0.10	 	& mag		& B\'ejar et~al. 2004b; Sherry et~al. 2008		\\ 
$[$Fe/H$]$		& --0.02$\pm$0.13	& --0.15:+0.13	 	& 		& Gonz\'alez-Hern\'andez et al. 2008			\\
Size, $r_{\rm max}$	& 30			& 20:40	 		& arcmin	& B\'ejar et~al. 2004a; Caballero 2008a			\\ 
Total mass, $\Sigma M$	& 275$^a$		& 150:225	 	& $M_\odot$	& Sherry et~al. 2004; Caballero 2007a			\\ 
Disc frequency$^b$	& $\sim$33		& 5:$>$50	 	& \%		& Caballero et~al. 2007; Luhman et~al. 2008		\\ %
\noalign{\smallskip}\hline\noalign{\smallskip}
\end{tabular}

$^a$ The value of $\Sigma M$ = 275\,$M_\odot$ is from Caballero (in~prep.).

$^b$ The disc frequency in $\sigma$~Orionis is mass-dependent and increases
towards lower masses.

\end{table}

The fourth brightest star in the Orion Belt, about 2\,mag fainter than the three
main stars, is {$\sigma$~Ori}. 
The star, which is actually the hierarchical multiple Trapezium-like stellar
system that illuminates the famous {Horsehead Nebula}, has taken a great
importance in the last decade. 
Its significance lies in the very early spectral type of the hottest component
($\sigma$~Ori~A, O9.5V) and in the homonymous star cluster that surrounds the
system (Garrison 1967).  
The $\sigma$~Orionis star cluster, re-discovered due to its large number of
X-ray emitters (Wolk 1996), contains one of the best known brown dwarf and
planetary-mass object populations (B\'ejar et~al. 1999, 2001; Zapatero Osorio
et~al. 2000, 2007; Gonz\'alez-Garc\'{\i}a et~al. 2006), and is an excellent
laboratory to study the evolution of X-ray emission, discs and angular momenta
(Reipurth et~al. 1998; Scholz \& Eisl\"offel 2004; Oliveira et~al. 2006;
Franciosini et~al. 2006; Hern\'andez et~al. 2007; Skinner et~al. 2008). 
Canonical, minimum and maximum values of main parameters of the cluster and {\em
some} key references are provided in Table~\ref{thetable}.
In this chapter, I~also described the work that the Canarias group had carried
out in $\sigma$~Orionis, with an emphasis on the discovery and characterisation
of S\,Ori~70, a mid-T-type object towards the cluster (Zapatero Osorio et~al.
2002b, 2008; Mart\'{\i}n \& Zapatero Osorio 2003; Burgasser et~al. 2004; Scholz
\& Jayawardhana 2008).
Finally, I~presented a compilation of cluster members with spectroscopic
confirmation that was the basis of two published catalogues of stars and brown
dwarfs in the $\sigma$~Orionis cluster (Caballero 2007a, 2008c).

\subsection{Multiobject spectroscopy in $\sigma$~Orionis: a bridge between the
stellar and substellar populations (Chapter 3)} 
\label{sec:chapter3}

\begin{figure*}[t]
\sidecaption[t]
\includegraphics[width=1.00\textwidth]{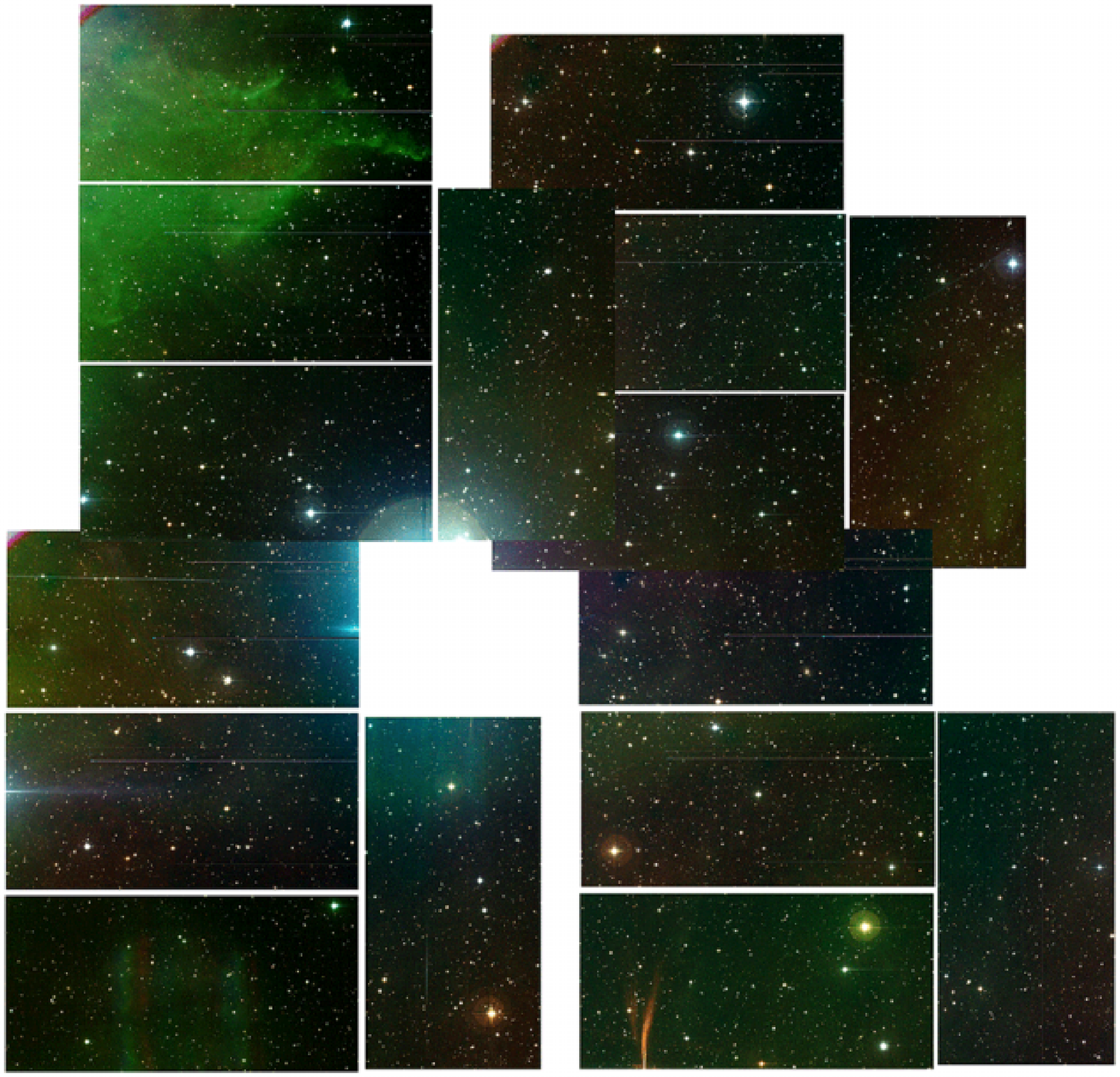}
\caption{False-colour mosaic of a region surrounding the $\sigma$~Orionis centre
(the bright star $\sigma$~Ori falls in the central gap).
The area corresponds to four pointings with the Wide Field Camera at the 2.5\,m
Isaac Newton Telescope.
Colours red, green and blue are for passbands $I$, $R$ and $V$, respectively.
Note the bright $R$-band (i.e. H$\alpha$) emission to the northeast; 
the nebulosity is associated with the Horsehead Nebula.} 
\label{thefigure}       
\end{figure*}

We used the Wide Field Fibre Optical Spectrograph instrument and the robot
positioner AutoFib2 (WYFFOS+AF2) at the 4.2\,m William Herschel Telescope to
aquire about 200 intermediate-resolution (R $\sim$ 8000) spectra of sources in
the direction of $\sigma$~Orionis.
We covered the wavelength range between 6400 and 6800\,\AA.
We compiled a list of 80 cluster members with WYFFOS+AF2 spectroscopy, based on
the presence of Li~{\sc i} $\lambda$6707.8\,{\AA} in absorption and H$\alpha$
$\lambda$6562.8\,{\AA} in emission (late and mid-type stars) or spectral type
determination (early-type stars).
About one half of the objects were spectroscopically studied there for the first
time.  
Using available data on the members, we investigated:
\begin{itemize}
\item the variation of the strength of the Li~{\sc i} line with spectral type
(from 0.05\,{\AA} in late F stars to 0.70\,{\AA} in intermediate M stars), time
and signal-to-noise ratio; 
\item the frequency of accretors according to the White \& Basri (2003)
criterion (46$^{+16}_{-13}$\,\% of K and M stars -- there might be a bias in the
input sample towards H$\alpha$ emitters) and the presence of asymmetries in the
profiles of the H$\alpha$ line;  
\item the existence of forbidden lines in emission ([N~{\sc ii}]
$\lambda\lambda$6548.0,6583.5\,{\AA}, [S~{\sc ii}]
$\lambda\lambda$6716.4,6730.8\,{\AA}); 
\item the widening of photospheric lines (of up to 100\,km\,s$^{-1}$) due to
fast rotation; 
\item the relationship between the $L'$- and $K_{\rm s}$-band flux excesses and
the spectroscopic features associated with accretion from protoplanetary discs;
\item the average of the radial velocity of the cluster members
(+30.2\,km\,s$^{-1}$) and the existence of radial velocity outliers (probably
due to unresolved close companions and contaminants of overlapping young stellar
populations in the Orion~Belt); 
\item and the frequency of X-ray emitters catalogued by {\em ROSAT} and {\em
ASCA} space observatories as a function of spectral type (the bulk of the K
stars are X-ray emitters). 
\end{itemize}

Our WYFFOS+AF2 data were used in the analysis of chemical abundances of
late-type pre-main sequence stars in $\sigma$~Orionis by Gonz\'alez-Hern\'andez
et~al. (2008), where we first determined the mean photospheric metallicity of
the cluster. 
Besides, we presented a new Herbig-Haro object candidate (a few arcseconds to
the southwest of the classical T~Tauri star Mayrit~609206)\footnote{Alternative
names to Mayrit objects listed in this work, in order of appearance --
Mayrit~609206: V505~Ori;
Mayrit~11238: $\sigma$~Ori~C;
Mayrit~13084: $\sigma$~Ori~D;
Mayrit~530005: S\,Ori~J053847.5--022711;
Mayrit~528005~AB: [W96]~4771--899;
Mayrit~3020~AB: $\sigma$~Ori~IRS1;
Mayrit~306125~AB: HD~37525;
Mayrit 208324: HD~29427;
Mayrit~1359077: HD~37686;
Mayrit~495216: S\,Ori~J053825.4--024241.} and
preliminary results on topics that have been developed afterwards, such as
radial distribution (Caballero 2008a) and wide binarity (Caballero~2008d).

\subsection{A new mini-search in the centre of $\sigma$~Orionis (Chapter 4)} 
\label{sec:chapter4}

Because of the intense brightness of the OB-type multiple star system
$\sigma$~Ori, the low-mass stellar and substellar populations close to the
centre of the very young $\sigma$~Orionis cluster was poorly know. 
I~presented an $IJHK_{\rm s}$ survey in the cluster centre, able to detect from
the massive early-type stars down to cluster members below the deuterium burning
mass limit.  
The near-infrared and optical data were complemented with X-ray imaging with the
{\em XMM-Newton} and {\em Chandra} space missions. 
Ten objects were found for the first time to display high-energy emission. 
Previously known stars with clear spectroscopic youth indicators and/or X-ray
emission defined a clear sequence in the $I$ vs. $I-K_{\rm s}$ diagram. 
I~ found six new candidate cluster members that followed this sequence. 
One of them, in the magnitude interval of the brown dwarfs in the cluster,
displayed X-ray emission and a very red $J-K_{\rm s}$ colour, indicative of a
disc\footnote{This object is actually an emission-line, Type~1, obscured quasar
at z = 0.2363$\pm$0.0005 (UCM0536--0239; Caballero et~al. 2002b).}. 
Other three low-mass stars have excesses in the $K_{\rm s}$ band as well. 
The frequency of X-ray emitters in the area is 80$\pm$20\,\%. 
The spatial density of stars is very high, of up to 1.6$\pm$0.1\,arcmin$^{-2}$. 
There was no indication of lower abundance of substellar objects in the cluster
centre. 
Finally, I also reported two cluster stars with X-ray emission located at only
8000--11000\,AU to $\sigma$~Ori~AFB, two sources with peculiar colours and an
object with X-ray emission and near-infrared magnitudes similar to those of
previously-known substellar objects in the cluster. 
({\bf A near-infrared/optical/X-ray survey in the centre of
$\sigma$~Orionis} -- Caballero~2007b)

\subsection{Multiplicity in $\sigma$~Orionis: adaptive optics in the near
infrared (Chapter 5)} 
\label{sec:chapter5}

Substellar objects, when companions to stars, are found in direct image at
distances larger than $\sim$40\,AU to the primaries (e.g. Nakajima et al. 1995,
Rebolo et al. 1998).  
While many multiple stellar systems and isolated substellar objects are found
in the $\sigma$~Orionis cluster, no brown dwarf or planetary-mass object at
projected physical separations from stellar members at less than about
10\,000\,AU has been {\em published} yet (but see the brown dwarf-exoplanet
system candidate in Caballero et~al. 2006b).
Through a pilot programme of near-infrared adaptive optic imaging with
Naomi+Ingrid at the William Herschel Telescope, we investigated the coronae
between 150 and 7000\,AU from six stellar cluster members.
The observed stars covered a wide range of spectral types, from O9.5V to K7.0.
Apart from the adaptive optic images, we used other near-infrared, optical and
X-ray data to derive the real astrophysical nature of the detected visual
companions.  
A total of 22 visual companions to the primary targets were detected in this
pencil-beam survey.  
Six sources showed blue optical-near-infrared colours for their magnitudes, and
they  did not match in any colour-magnitude diagram of the cluster.
There is not enough information to derive the nature of other five sources
(including a faint object 2\,arcsec northeast of Mayrit~11238; see Bouy et~al.
2008).  
Eleven objects remained as cluster member candidates according to their
magnitudes and colours:
($i$) three of them were previously known cluster members:
Mayrit~11238, Mayrit 13084 (surrounding $\sigma$~Ori~AFB) and
Mayrit~530005 (close to Mayrit~528005~AB);
($ii$) one is the near-infrared counterpart of the mid-infrared and radio
source Mayrit~3020~AB, a dust cloud next to $\sigma$~Ori~AFB discovered by
van~Loon \& Oliveira (2003). 
The object was also detected in {\em Chandra} archive images taken with the
HRC-I instrument (this result was advanced in Caballero 2005; see again Bouy
et~al. 2008); 
($iii$) one of the Mayrit~306125~AB companions seemed to be a pre-main sequence
photometric candidate star catalogued by Wolk (1996); 
($iv$) two bright objects were the previously unknown secondaries of the
Mayrit 306125~AB and Mayrit~528005~AB close binary systems, at angular
separations of 0.45$\pm$0.04 and 0.40$\pm$0.08\,arcsec, respectively; and 
($v$) the four remaining objects were visual companions to $\sigma$~Ori AFB (1),
Mayrit 208324 (2) and Mayrit~1359077 (1) at separations from 5.5 to 19.0\,arcsec. 
A few of them display features of youth (e.g. discs).
Even if their common spatial velocities are measured in the future, it is not
known whether the systems will survive the gravitational field of the young
cluster.

\subsection{The mass function down to the planetary domain: the ``Anaga'' survey
(Chapter 6)} 
\label{sec:chapter6}

We investigated the mass function in the substellar domain of the
$\sigma$~Orionis open cluster down to a few Jupiter masses.   
We performed a deep $IJ$-band search with Isaac at the 8.2\,m Very Large
Telescope UT1 and the Wide Field Camera at the Isaac Newton Telescope, covering
an area of 790\,arcmin$^2$ close to the cluster centre.   
This survey was complemented with an infrared follow-up in the $HK_{\rm s}$- and
3.6-8.0\,$\mu$m-bands with IRAC at the {\em Spitzer Space Telescope}, CFHT-IR at
the 3.6\,m Canada-France-Hawai'i Telescope, Omega-2000 at the 3.5\,m Calar Alto
Teleskop and CAIN-II at the 1.5\,m Telescopio Carlos S\'anchez. 
Using colour-magnitude diagrams, we selected 49 candidate cluster members
in the magnitude interval 16.1\,mag $< I <$ 23.0\,mag.  
Accounting for flux excesses at 8.0\,$\mu$m and previously known spectral
features of youth, we identified 30 objects as bona fide cluster members.
Four were first identified from our optical/near-infrared data. 
Eleven had most probable masses below the deuterium burning limit, which we
therefore classified as candidate planetary-mass objects. 
The slope of the substellar mass spectrum ($\Delta N / \Delta M \approx a
M^{-\alpha}$) in the mass interval 0.11\,$M_\odot$  $< M <$ 0.006\,$M_\odot$ is
$\alpha$ = +0.6$\pm$0.2. 
Any mass limit to formation via opacity-limited fragmentation must lie below
0.006\,$M_\odot$. 
The frequency of $\sigma$~Orionis brown dwarfs with circumsubstellar discs is
47$\pm$9\,\%.  
The continuity in the mass function and in the frequency of discs suggests that
very low-mass stars and substellar objects, even below the deuterium-burning
mass limit, share the same formation mechanism.
Besides, the technique used for calculating in detail the back- and foreground
contamination by field dwarfs of very late spectral types (intermediate and late
M, L, and T) that we presented in this chapter was developed by Caballero
et~al.~(2008a) with the adoption of the latest models from the literature.
  ({\bf The substellar mass function in $\sigma$~Orionis. II. Optical,
near-infrared and IRAC/{\em Spitzer} photometry of young cluster brown dwarfs
and planetary-mass objects} -- Caballero et~al.~2007)

\section{Activity and meteorology in ultracool objects: discs and atmospheres
(Part III)} 
\label{sec:part3}

\subsection{Photometric variability of young brown dwarfs in $\sigma$~Orionis
(Chapter 7)} 
\label{sec:chapter7}

We carried out multi-epoch, time-series differential $I$-band photometry of a
large sample of objects in the south-east region of the $\sigma$~Orionis open
cluster.  
A field of $\sim$1000\,arcmin$^2$ was monitored with the Wide Field Camera at
the Isaac Newton Telescope during four nights over a period of two years.  
Using this dataset, we studied the photometric variability of twenty-eight brown
dwarf cluster candidates with masses ranging from the stellar-substellar
boundary down to the planetary-mass domain.  
We found that about 50\,\% of the sample showed photometric variability on
timescales from less than one hour to several days and years. 
The amplitudes of the $I$-band light curves ranged from less than 0.01 up to
$\sim$0.4\,mag. 
A correlation between the near-infrared excess in the $K_{\rm s}$ band, strong
H$\alpha$ emission and large-amplitude photometric variation was observed. 
We briefly discussed how these results may fit the different scenarios
proposed to explain the variability of cool and ultracool dwarfs (i.e. magnetic
spots, patchy obscuration by dust clouds, surrounding accretion discs and
binarity). 
Additionally, we determined tentative rotational periods in the range 3 to 40\,h
for three objects with masses around 60\,$M_{\rm Jup}$, and the rotational
velocity of 14$\pm$4\,km\,s$^{-1}$ for one of them.
The shortest periods can be explained by pulsational instability excited by
central deuterium burning during the initial phases of evolution of young brown
dwarfs.
({\bf Photometric variability of young brown dwarfs in the
$\sigma$~Orionis open cluster} -- Caballero et~al. 2004)

\subsection{S\,Ori~J053825.4--024241: a classical T~Tauri-like object at the
substellar boundary (Chapter 8)} 
\label{sec:chapter8}

We presented a spectrophotometric analysis of Mayrit~495216
(S\,Ori~J053825.4--024241), a candidate member close to the substellar boundary
of the $\sigma$~Orionis cluster.  
Our optical and near-infrared photometry and low-resolution spectroscopy
indicated that Mayrit~495216 is a likely cluster member with a mass
estimated from evolutionary models at $0.06^{+0.07}_{-0.02}$\,$M_\odot$, which
made the object a probable brown dwarf.  
The radial velocity of Mayrit~495216 was similar to the cluster
systemic velocity.  
This target, which we classified as an M\,6.0$\pm$1.0 low-gravity object,
showed excess emission in the near-infrared and anomalously strong photometric
variability for its type (from the blue to the $J$ band), suggesting the
presence of a surrounding disc. 
The optical spectroscopic observations showed a continuum excess at short
wavelengths and a persistent and resolved H$\alpha$ emission (pseudo-equivalent
width of about --250\,\AA) in addition to the presence of other forbidden and
permitted emission lines, which we interpret as indicating accretion from the
disc and possibly mass loss. 
We concluded that despite the low mass of Mayrit~495216, this object
exhibits some of the properties typical of active classical T Tauri stars.
[{\em Facilities}: LRIS at the 10.0\,m Keck I Telescope, ALFOSC at the 2.6\,m
Nordic Optical Telescope, CAIN-II at the Telescopio Carlos S\'anchez, ESACCD at
the 1.0\,m European Space Agency Optical Ground Station, CCD\#1 at the 0.8\,m
Telescopio IAC-80]
({\bf S\,Ori~J053825.4--024241: a classical T~Tauri-like object at
the substellar boundary} -- Caballero et~al. 2006a)

\section{Very low-mass companions to young stars and ultracool dwarfs in the
solar nighbourhood (Part IV)} 
\label{sec:part4}

\subsection{A search for very low-mass objects around nearby young stars (Chapter
9)} 
\label{sec:chapter9}

There is a strong competition for searching for and characterising 
resolved brown dwarfs and exoplanets in orbit to neighbour stars, being the
ultimate goal of many astronomers the imaging of {\em exoearths} in
solar-like systems.
Given the overluminosity of  very low mass objects during their contraction
phase, most high spatial-resolution, photo(astro)metric searches have
tended to explore nearby ($d <$ 100\,pc), young ($\tau \sim$ 10--600\,Ma) stars
(e.g. Neuh\"auser et~al. 2003; McCarthy \& Zuckerman 2004; Metchev \&
Hillenbrand 2004; Masciadri et~al. 2005; Lowrance et~al. 2005; Biller et~al.
2007; Lafreni\`ere et~al. 2007; Carson et~al. 2008). 
Following this idea, we imaged 51 stellar systems with features of youth
(lithium, chromospheric activity, X-ray emission, membership in moving group)
with the NICMOS instrument and the coronograph at the {\em Hubble Space
Telescope} and with near infrared adaptive optics systems attached at 4 m-class
telescopes: 
Alfa+Omega-Cass at the 3.5\,m Calar Alto Teleskop, AdOpt@TNG+NICS at the
3.6\,m Telescopio Nazionale Galileo and, especially, Naomi+Ingrid at the William
Herschel Telescope. 
High resolution images were complemented with wide-field searches with CAIN-II
at the Telescopio Carlos S\'anchez and other near-infrared and optical
instruments.
Of the 51 investigated systems, 32 (44) are 100\,Ma (600\,Ma) old or younger.
The survey was designed to detect all brown dwarfs at projected physical
separations $\Delta >$ 50\,AU and all exoplanets with $M_2 >$ 0.008\,$M_{\odot}$
at $\Delta >$ 100\,AU.
However, we did not detect any new substellar object.
Complementing our results with those in the literature, the frequency of
substellar objects with $M_2 >$ 0.008\,$M_{\odot}$ is less than 2\,\% at any
distance interval.
Besides, we discovered three, possibly four, new stellar companions ($M_2 \sim$
0.35--0.80\,$M_{\odot}$) and measured accurate astrometry ($\rho$, $\theta$) of
a dozen young, late-type, close binaries.
From a personal point of view, the search was characterised by our ``bad luck'':
AB~Dor~C (Close et~al. 2005) was below the NICMOS coronographic mask; HN~Peg~B
(Luhman et~al. 2007) was out of the Naomi+Ingrid field of view and had no
optical images to complement with our wide-field near-infrared ones; the
individual exposure time for the binary HD~160934~AC (Hormuth et~al. 2007) was
too long and we could not resolve it, etc.

\subsection{Multiplicity of L dwarfs: binarity and habitable planets (Chapter
10)}  
\label{sec:chapter10}

On the one hand, stars do have planets. 
The least massive exoplanets found to date, with a few Earth masses
($M_\oplus$), orbit low-mass, M-type stars. 
On the other hand, the most massive moons in the Solar System, with up to
0.025\,$M_\oplus$, orbit giant planets.
Brown dwarfs, with masses in between the least massive stars and the most
massive giant planets, also have planets in wide (Chauvin et~al. 2004) and close
orbits (Joergens \& M\"uller 2007).
Besides, the frequency of (protoplanetary) discs in brown dwarfs is comparable,
or even larger, than in stars (see, for example, Caballero et~al. 2007;
Section~\ref{sec:chapter6}).
Therefore, it is natural to hypothesise the existence of terrestrial planets
surrounding very low-mass stars and brown dwarfs with spectral types L and T.
Because of their intrinsic dimmness, the habitable zones are very close to the
Roche limit of the central objects.
Such kind of systems can be detected with {\em current} technology.
In this chapter, I~showed preliminary results on photometric monitoring at
medium-size telescopes to search for transits (Caballero \& Rebolo 2002; Blake
et~al. 2008), that resulted in a search for variability in brown dwarf
atmospheres and for wide faint companions (Goldman 2003; Caballero et~al. 2003;
Goldman et~al. 2008), and detailed a methodology for detecting {\em exoearths}
in habitable zones around nearby L (and T) dwarfs with high-resolution
near-infrared spectrographs (e.g. Nahual at the 10.4\,m Gran Telescopio
Canarias).

\section{Conclusions, appendices and bibliography (Part V)} 
\label{sec:3}

\subsection{Summary (Chapter 11)} 
\label{sec:chapter11}

As a corollary of my thesis, the frequency of substellar companions is low,
whether around nearby stars in the field or whether close to very young stars in
the $\sigma$~Orionis cluster. 
However, isolated brown dwarfs and planetary-mass objects in clusters represent
a significative fraction of the total number of objects (but not of the total
mass).
The similarity in spatial distribution and the continuity in the rising mass
spectrum and in the frequency of discs suggest that very low-mass stars and
substellar objects, even below the deuterium-burning mass limit, share the same
formation mechanism.   
If they formed in protoplanetary discs by gravitational instabilities, a very
efficient ejection mechanism would be necessary during the first few million 
years. 
Thus, the isolated planetary-mass objects that we find free-floating in clusters
likely formed from turbulent fragmentation in the primigenious gas cloud. 

This thesis is a ``full stop, new sentence'' in the quotation {\em Smaller,
Fainter, Cooler} (in humorous contraposition to {\em Bigger, Stronger, Faster}) 
of the brown dwarf and exoplanet searches.

\begin{acknowledgement}
I thank R. Rebolo and V.~J.~S. B\'ejar for helpful comments an innumerable
individuals and groups for their friendship and assistance during my PhD.
Especial gratitude is for {\em uKi} and {\sc m4m4}.
Most of the thesis research was conducted during my residence at the Instituto 
de Astrof\'{i}sica de Canarias.
Partial financial support was provided by a number of projects of the Spanish
Ministerio Educaci\'on y Ciencia, Ministerio de Ciencia y Tecnolog\'{\i}a,
Comunidad Aut\'onoma de Madrid, Universidad Complutense de Madrid, Spanish
Virtual Observatory, and European Social Fund.
\end{acknowledgement}

\end{document}